\documentclass[10pt, onecolumn, conference]{IEEEtran}

\usepackage{graphicx,url,tabularx,booktabs}
\usepackage[utf8]{inputenc}
\usepackage{comment}
\usepackage{ifthen}
\usepackage{tcolorbox}
\usepackage{orcidlink}
\usepackage{amssymb}
\usepackage{pifont}
\usepackage[table]{xcolor}
\usepackage{listings}
\usepackage{enumitem}
\usepackage{amsmath}
\title{A Reproducible Semantic Benchmark for Multivendor DSM-to-CLI Translation}

\author{
\IEEEauthorblockN{
    Jerônimo Menezes\IEEEauthorrefmark{1}\IEEEauthorrefmark{2},
    Leonardo Bitzki\IEEEauthorrefmark{1}\IEEEauthorrefmark{2},
    Diego Kreutz\IEEEauthorrefmark{2},\\
    Gefte Almeida\IEEEauthorrefmark{2},
    Marcio Pohlmann\IEEEauthorrefmark{1}\IEEEauthorrefmark{2},
    Rodrigo Mansilha\IEEEauthorrefmark{2}
}
\IEEEauthorblockA{\IEEEauthorrefmark{1}Federal University of Rio Grande do Sul (UFRGS)}
\IEEEauthorblockA{\IEEEauthorrefmark{2}AI Horizon Labs and PPGES -- Federal University of Pampa (UNIPAMPA)}
}

\begin{document}

\maketitle

\begin{abstract}
Translating high-level network intents into correct multivendor configurations remains a central challenge in network automation, as syntactically valid outputs may still violate the intended operational state. Despite recent advances in Large Language Models (LLMs), the field still lacks reproducible semantic benchmarks for rigorous cross-vendor evaluation. This paper presents a reproducible DSM-to-CLI semantic benchmark covering five cloud LLMs, three vendors, five representative use cases, and ten repeated runs per experimental cell under fixed judges and an explicit failure taxonomy. Our results show that semantic quality and operational reliability are orthogonal, vendor effects dominate use-case effects, and repeated-run dispersion strongly predicts vote instability, with Huawei VRP exposing failure modes hidden by aggregate metrics. These findings demonstrate that multivendor, repeated-execution semantic benchmarks are essential for scientifically rigorous comparison of LLM-based network configuration systems.
\end{abstract}

\section{Introduction}
\label{sec:introduction}

Network configuration in heterogeneous, multivendor environments remains one of the most persistent operational challenges in modern network management. Differences in command syntax, vendor-specific semantics, feature support, and configuration abstractions make automation pipelines difficult to generalize and highly susceptible to silent failures~\cite{boateng2025survey,hong2025comprehensive,llm-powered}. In practice, syntactically valid configurations do not necessarily implement the intended operational behavior, particularly when equivalent policies must be mapped across distinct vendor ecosystems. This gap between syntactic validity and semantic correctness remains a central obstacle to robust network automation~\cite{wei_inta_2025,tageldien2025large}.

Recent advances in Large Language Models (LLMs) have renewed interest in intent-driven network automation, with multiple studies exploring the translation of natural language intents or structured specifications into device-specific configurations~\cite{lin2025agentic,net2dllm2025,netconfeval2025,peeringllmbench2025}. These results demonstrate the practical feasibility of LLM-based translation pipelines. However, as highlighted by recent surveys~\cite{boateng2025survey,hong2025comprehensive}, the current state of the art remains methodologically immature. Most existing approaches still frame the problem primarily as text generation, emphasizing prompt engineering, lexical similarity, or execution success, while providing limited guarantees regarding semantic fidelity, reproducibility, stability across executions, and comparability across models and vendors.

This limitation is particularly critical in Intent-Based Networking (IBN), where the objective is not merely to generate plausible commands, but to ensure that vendor-specific configurations faithfully realize the intended network state~\cite{wei_inta_2025}. In such settings, evaluation based on syntactic plausibility, string matching, or isolated functional probes is fundamentally insufficient. Multiple command sequences may be syntactically different yet semantically equivalent, while seemingly correct outputs may still violate the declared intent. Consequently, rigorous progress in this domain requires semantically grounded benchmarks, independent verification pipelines, repeated-run analysis, and controlled multivendor experimentation~\cite{tageldien2025large,11349540}.

In previous work, we introduced \textit{dsm2cli}\footnote{\url{https://github.com/net2d-community/dsm2cli}}, a verifiable and observable pipeline for translating structured network intents, expressed as Desired State Models (DSMs), into multivendor CLI. The framework explicitly separates translation from semantic verification, employs independent evaluators to assess adherence between the DSM and the generated configuration, and records structured artifacts including verdicts, voting traces, telemetry, and failure evidence. That work demonstrated that semantic failures often remain hidden in generation-only workflows and that explicit verification significantly improves interpretability and traceability.

Building on this foundation, this paper moves beyond system design toward benchmark methodology by introducing \textit{dsm2cli-bench}\footnote{\url{https://github.com/net2d-community/dsm2cli-bench}}, an extensive and reproducible evaluation framework for DSM-to-CLI translation. Our goal is to define and apply a rigorous semantic benchmark for the systematic comparison of cloud LLMs in multivendor configuration generation tasks. The proposed benchmark spans three vendors, five representative use cases, fixed judge models, repeated executions, an explicit failure taxonomy, and a pre-specified statistical analysis protocol.
Rather than asking only which model achieves the highest average accuracy, we perform a multidimensional evaluation encompassing semantic correctness, execution stability, inter-judge agreement, vendor sensitivity, latency, and token efficiency under a unified and fully reproducible experimental framework.

The main contributions of this paper are fourfold: first, we design a reproducible semantic benchmark for DSM-to-CLI translation in multivendor environments; second, we introduce a controlled experimental protocol that combines structured DSM inputs, fixed judges, repeated executions, and an explicit failure taxonomy; third, we provide a comparative evaluation of five cloud LLM translators across Cisco NX-OS, Huawei VRP, and Arista EOS; and finally, we present a multidimensional analysis encompassing semantic correctness, execution stability, inter-evaluator agreement, latency, and token usage.

% The remainder of this paper is organized as follows. Section~\ref{sec:related_work} discusses related work. Section~\ref{sec:methodology} presents the benchmark design and methodology. Section~\ref{sec:results} reports the experimental results. Section~\ref{sec:discussion} discusses the main findings. Section~\ref{sec:threats} presents threats to validity, and Section~\ref{sec:conclusion} concludes the paper.

\section{Related Work}
\label{sec:related_work}

Recent studies and surveys have shown a clear convergence toward the use of Large Language Models (LLMs) for automating network management and configuration tasks, particularly through pipelines that translate natural language intents into vendor-specific commands, structured representations, or executable workflows~\cite{long2025survey,boateng2025survey,hong2025comprehensive,liu2024large}.
However, the state of the art remains scientifically heterogeneous and methodologically immature across the dimensions summarized in Table~\ref{tab:soa_comparison}.

\begin{table*}[!htp]
\centering
\caption{Comparative analysis of LLM-based network configuration approaches.}
\label{tab:soa_comparison}
\renewcommand{\arraystretch}{1.3}
\setlength{\tabcolsep}{4pt}
\resizebox{\textwidth}{!}{
\begin{tabular}{lccccc}
\hline
\textbf{Paper} & \textbf{Input $\rightarrow$ Output} & \textbf{Multivendor} & \textbf{Semantic Evaluation} & \textbf{Repeated Runs} & \textbf{Benchmark Protocol} \\
\hline
Lin et al. & NL $\rightarrow$ YANG $\rightarrow$ CLI & Partial & No & Limited (3 trials) & No \\
Net2d-LLM & DSM $\rightarrow$ CLI & No & No & No & No \\
NetConfEval & NL $\rightarrow$ Spec/API/CLI & No & Lexical only & Limited & Partial \\
PeeringLLM-Bench & NL $\rightarrow$ CLI & Yes & Binary/functional & Yes & Yes \\
NetLLMBench & NL $\rightarrow$ Execution & No & No & No & No \\
INTA & CLI $\rightarrow$ CLI & Limited & No & No & No \\
\textbf{dsm2cli (ours)} & \textbf{DSM $\rightarrow$ CLI} & \textbf{Yes} & \textbf{Yes (formal)} & \textbf{Yes} & \textbf{Yes (reproducible)} \\
\hline
\end{tabular}
}
\end{table*}

The analyzed literature shows that most existing approaches focus on direct translation pipelines, typically NL$\rightarrow$CLI or NL$\rightarrow$IR$\rightarrow$CLI, as in Lin et al.~\cite{lin2025agentic}, Net2d-LLM~\cite{net2dllm2025}, and NetConfEval~\cite{netconfeval2025}. While these studies demonstrate the feasibility of LLM-based intent-driven configuration, they largely rely on prompt engineering, single-pass inference, or loosely structured workflows without a formally defined intermediate representation, limiting interpretability, reproducibility, and systematic error analysis.

A second major limitation concerns vendor coverage. Most prior studies operate in single-vendor or pseudo-multivendor settings, where heterogeneity is only partially represented. Even works explicitly targeting multivendor environments, such as PeeringLLMINTA-Bench~\cite{peeringllmbench2025} and INTA~\cite{inta2025}, evaluate only a narrow subset of vendors or protocols, typically centered on BGP peering or CLI translation. Recent surveys further highlight broad multivendor validation as one of the main open challenges in the field~\cite{hong2025comprehensive,boateng2025survey}.

More critically, rigorous semantic evaluation remains largely absent. Most existing studies rely on textual similarity, syntactic validation, execution-based proxies, or human judgment~\cite{netconfeval2025,netllmbench2025}, which fail to capture whether the generated configuration faithfully preserves the intended network state. This issue is particularly severe in multivendor settings, where distinct command sequences may implement the same policy semantics.

Another recurring weakness is the lack of repeated-run and statistical stability analysis. Given the stochastic nature of LLM outputs, single-run experiments or limited trial sets provide insufficient scientific evidence. Most prior work does not quantify variance, reproducibility, or output consistency across executions, weakening the reliability and generalizability of the reported results~\cite{boateng2025survey,hong2025comprehensive,long2025survey}.

In contrast, our work advances the state of the art through a structured and reproducible DSM$\rightarrow$CLI benchmark that combines a formally defined intermediate representation, rigorous semantic evaluation with fixed judges, repeated-run stability analysis, and systematic multivendor comparison, enabling a substantially more rigorous assessment of LLM-based network configuration automation.

\section{Benchmark Design and Methodology}
\label{sec:methodology}

This section presents the benchmark design, experimental protocol, evaluation metrics, and statistical procedures adopted to systematically assess DSM-to-CLI semantic translation in multivendor environments.

\subsection{Research Questions and Hypotheses}
\label{sec:rqs_hypotheses}

The benchmark is designed to investigate three fundamental dimensions of structured-intent translation into vendor-specific CLI: cross-vendor variability, scenario-dependent semantic difficulty, and execution stability under repeated runs. Specifically, we hypothesize that performance differences will be more pronounced for Huawei VRP than for Cisco NX-OS, while Arista EOS is expected to behave more similarly to Cisco. We further expect SVI-based scenarios to exhibit lower semantic success rates and lower evaluator consensus than Layer 2 scenarios. Finally, we hypothesize that greater dispersion in binary outcomes across repetitions is associated with lower vote stability, independently of mean correctness.

\subsection{Experimental Harness}
\label{sec:harness}

The benchmark adopts \textit{dsm2cli} as a controlled experimental harness. For each execution, the system receives a DSM, device metadata, and a selected translator model, producing vendor-specific CLI that is subsequently evaluated by a fixed panel of three independent judges. Their individual votes are aggregated into a final semantic verdict, while all intermediate artifacts are persistently recorded.

This design enforces strict separation between generation and semantic verification, ensuring a stable input--output contract across all translators. It also preserves structured artifacts, including generated CLI, individual judge votes, final verdicts, failure labels, and execution telemetry. Throughout all experiments, the DSM schema, judge panel, failure taxonomy, and telemetry pipeline remain fixed, with the translator model as the primary experimental factor.

\subsection{Experimental Matrix}
\label{sec:vendors_usecases}

The experimental matrix spans three vendors, Cisco NX-OS, Huawei VRP, and Arista EOS, across five representative use cases: UC1 (L2 access), UC2 (L2 trunk with native VLAN), UC3 (SVI IPv4), UC4 (SVI IPv6), and UC5 (dual-stack SVI).

The DSM is treated as the complete normative specification. No prior device state is assumed, and vendor defaults are considered valid only when explicitly encoded in the DSM. This yields 15 scenarios per translator (3 vendors $\times$ 5 use cases). Each scenario is executed 10 times, resulting in 150 runs per translator and 750 executions overall.

\subsection{Translators and Judges}
\label{sec:models_judges}

Five cloud-based translators are evaluated: \texttt{gpt-5}, \texttt{claude-opus-4-6}, \texttt{gemini-2.5-pro}, \texttt{deepseek-chat}, and \texttt{grok-4-1-fast-reasoning}, covering distinct provider families under a common semantic translation protocol.

Semantic evaluation is performed by a fixed panel of three judges: \texttt{gpt-5.4-mini}, \texttt{claude-haiku-4-5}, and \texttt{gemini-2.5-flash-lite}. This configuration explicitly avoids self-evaluation and minimizes evaluator drift, thereby ensuring direct comparability across translators, vendors, and repeated executions.

\subsection{Execution Protocol}
\label{sec:protocol}

Each execution is defined by the tuple (translator, vendor, use case, repetition). Experiments are organized into 10 rounds, in which all 75 experimental cells (5 translators $\times$ 3 vendors $\times$ 5 use cases) are executed once per round in randomized order. This procedure mitigates temporal effects and provider-side bias.

Translator temperature is fixed at 0.2, while judge temperature is set to 0.0 to maximize evaluation determinism. Maximum output lengths are capped at 768 tokens for translators and 1536 tokens for judges. A timeout threshold of 300 seconds is enforced, with up to two automatic retries for technical failures. Runs that remain unsuccessful after retries are classified as pipeline errors.

\subsection{Metrics and Failure Taxonomy}
\label{sec:metrics_taxonomy}

The benchmark collects three classes of metrics: semantic quality, consistency, and performance. Semantic quality is quantified through success rate, mean favorable votes, semantic failure rate, and pipeline error rate. Consistency is measured at the cell level using vote dispersion and the Vote Stability Index (VSI), defined as the proportion of repetitions assigned to the modal vote category. Performance is assessed through mean latency, 95th-percentile latency, and token usage. Semantic failures are classified as \textbf{omission}, \textbf{divergence}, and \textbf{structural}, whereas pipeline failures are categorized as \textbf{timeout}, \textbf{provider error}, \textbf{invalid output}, \textbf{parsing failure}, and \textbf{empty response}.

\subsection{Statistical Analysis}
\label{sec:statistical_analysis}

The statistical analysis is fully specified \textit{a priori}. Hypotheses H1 and H2 are evaluated through planned proportion contrasts using Wilson confidence intervals and two-proportion z-tests. Hypothesis H3 is assessed at the cell level through ordinary least squares regression:
\[
\text{VSI} \sim \texttt{sd\_outcome}
\]
where \texttt{sd\_outcome} denotes the standard deviation of the binary correct/not-correct outcome across the 10 repetitions of each cell. Mean accuracy is intentionally excluded from this model to avoid structural coupling between correctness and stability. Inter-judge agreement is quantified using Fleiss' Kappa and percent agreement, stratified by vendor and use case.

\subsection{Pilot Calibration}
\label{sec:pilot_calibration}

A pilot phase was conducted to validate the benchmark implementation and led to two important protocol refinements: improved semantic equivalence handling for UC2 under vendor-specific CLI syntax and explicit \texttt{pipeline\_error\_type} labeling to improve observability of technical failures.

This refinement proved critical in the final experiment, where 17 HTTP~400 errors from \texttt{claude-haiku-4-5}, initially misclassified as structural translator failures, were subsequently reclassified as judge-side faults. The UC2 calibration also strengthened benchmark validity, which is reflected in UC2 emerging as the best-performing and highest-agreement scenario in the final dataset.

\section{Experimental Results}
\label{sec:results}

This section presents the final benchmark results aggregated across all ten experimental rounds, totaling 750 executions. The analyses that follow draw on distinct subsets of these runs depending on the specific metric under consideration, as summarized in Table~\ref{tab:outcome_summary}. Translator correctness metrics are computed over the full set of 750 executions, providing a comprehensive view of raw performance, while semantic quality metrics are evaluated exclusively on the 667 valid runs that successfully completed without pipeline errors or judge faults to ensure that semantic assessments reflect genuine translator capabilities rather than technical failures. Inter-judge agreement is further restricted to a 484-run subset derived from these valid runs, corresponding to executions where all three judges produced evaluable outputs, enabling consistent measurement of evaluator consensus. All raw data generated throughout the experiment, including individual CLI outputs, judge verdicts, telemetry logs, and failure classifications, are publicly available in the project's GitHub repository\footnote{\url{https://github.com/net2d-community/wiarc2026-experiment}}, ensuring full reproducibility and independent verification of the results.

For the agreement analysis, we exclude 183 executions in which one judge classified the output as \texttt{structural} while the other two judges marked it as correct. These cases reflect categorical disagreement rather than technical failure and are therefore excluded from the Fleiss' Kappa computation to focus the analysis on binary correct-versus-incorrect divergence.

\begin{table}[ht]
\centering
\small
\caption{Execution outcome summary and analysis subsets.}
\label{tab:outcome_summary}
\setlength{\tabcolsep}{5pt}
\renewcommand{\arraystretch}{1.3}
\begin{tabular}{lrrr}
\toprule
\textbf{Outcome} & \textbf{Raw} & \textbf{Final} & \textbf{Rate} \\
\midrule
Correct (semantic)  & 655 & 655 & 87.3\% \\
Semantic incorrect  &  29 &  12 &  1.6\% \\
Judge fault (infra) &   0 &  17 &  2.3\% \\
Pipeline error      &  66 &  66 &  8.8\% \\
\midrule
Total               & 750 & 750 & 100.0\% \\
\bottomrule
\end{tabular}
\end{table}

\subsection{Infrastructure Events and Outcome Classification}
\label{sec:infra_events}

Before semantic analysis, two infrastructure-level events must be accounted for. First, starting in round~7, \texttt{claude-opus-4-6} experienced a sustained provider-side outage, resulting in 50 \texttt{provider\_error} outcomes out of 150 scheduled runs. Importantly, all 100 executions completed before the outage were semantically correct.

Second, 17 runs were affected by HTTP~400 credential errors in the judge model \texttt{claude-haiku-4-5}. These cases were initially logged as structural translator failures because the pipeline stored the judge error message as a semantic finding. Artifact inspection confirmed that the fault originated in the evaluation layer rather than in the generated CLI, and these runs were therefore reclassified as \emph{judge faults}.

\subsection{Overall Results}
\label{sec:overall_results}

After reclassifying judge faults, the benchmark results comprise 655 correct executions, 12 genuine semantic failures, 17 judge faults, and 66 pipeline errors across 750 runs. Excluding infrastructure-related events, the semantic success rate over valid judged runs reaches 98.2\%.

The 12 genuine semantic failures consist of 11 omissions and 1 divergence, with no translator-generated structural errors. This suggests that semantic failures arise predominantly from underspecification of the intended configuration rather than explicit contradiction of the DSM.

\subsection{Results by Translator}
\label{sec:results_translator}

At the translator level, \texttt{grok-4-1-fast-reasoning} achieves the best end-to-end correctness, with 149 correct executions out of 150 and no infrastructure-related disruptions (Table~\ref{tab:translator_results}). In contrast, \texttt{claude-opus-4-6} achieves perfect semantic correctness on all valid judged runs, but its end-to-end performance is substantially reduced by the provider-side outage that affected 50 executions. \texttt{deepseek-chat} performs strongly on Cisco and Arista, but its overall score is degraded by both semantic failures and pipeline errors concentrated in Huawei scenarios.

\begin{figure}[!htp]
\centering
\includegraphics[width=0.8\linewidth]{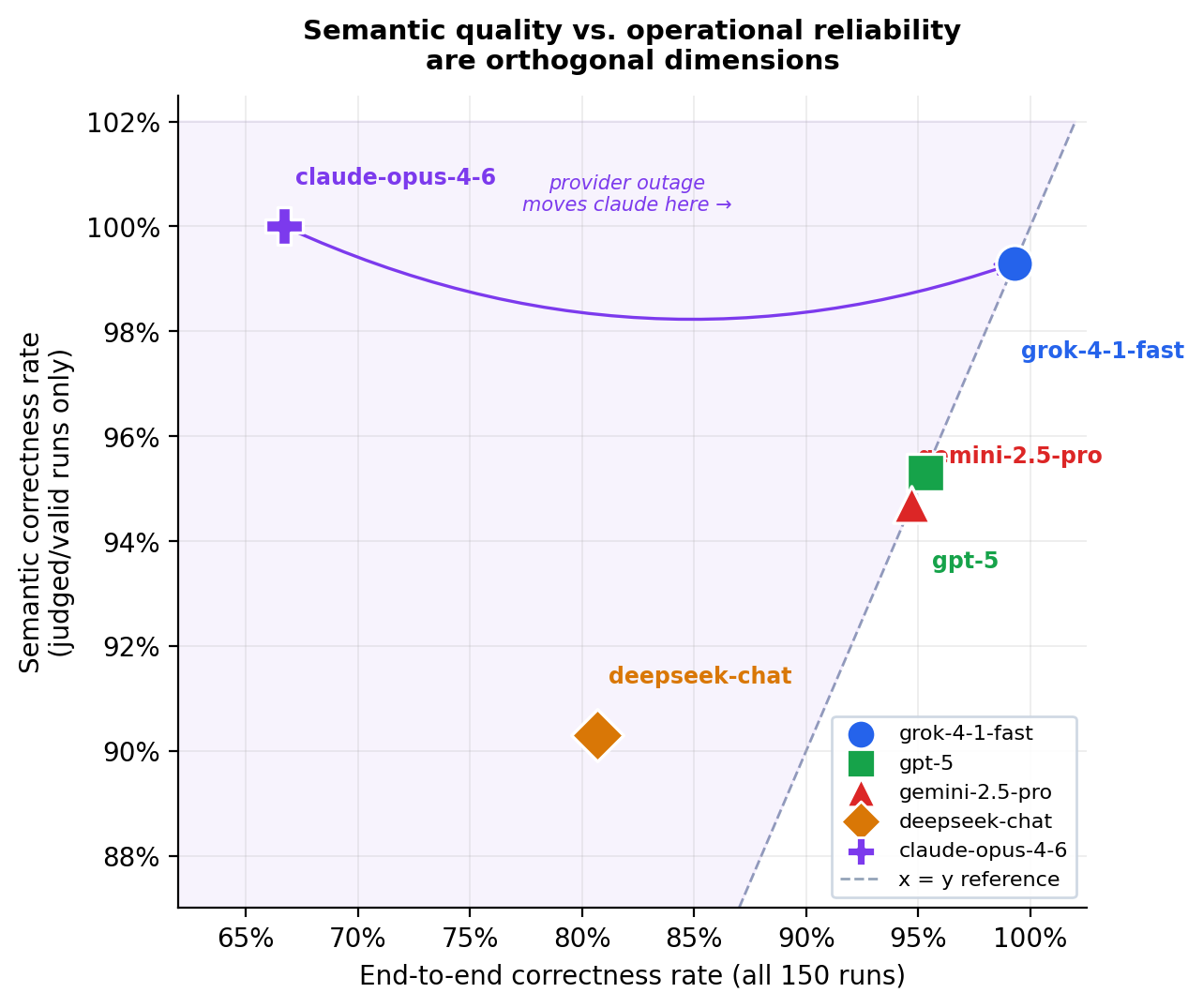}
\caption{Semantic versus end-to-end correctness by translator.}
\label{fig:quality_vs_reliability}
\end{figure}

These results highlight that semantic quality and operational reliability are orthogonal dimensions. In particular, \texttt{claude-opus-4-6} remains semantically perfect despite ranking lower in the end-to-end view, reinforcing the need to report both metrics separately, as illustrated in Figure~\ref{fig:quality_vs_reliability}.

\begin{table}[!htp]
\centering
\small
\caption{Translator-level correctness, semantic failures, and pipeline errors.}
\label{tab:translator_results}
\renewcommand{\arraystretch}{1.3}
\setlength{\tabcolsep}{4pt}
\begin{tabular}{lrrrrrr}
\toprule
\textbf{Translator} &
\textbf{Corr./Tot.} &
\textbf{Rate} &
\textbf{Sem.} &
\textbf{Pipe.} &
\textbf{$\mu$ votes} &
\textbf{$\sigma$ votes} \\
\midrule
\texttt{grok-4-1-fast-reasoning} & 149/150 & 99.3\%          & 1  & 0  & 2.660 & 0.489 \\
\texttt{gpt-5}                   & 143/150 & 95.3\%          & 3  & 0  & 2.507 & 0.588 \\
\texttt{gemini-2.5-pro}          & 142/150 & 94.7\%          & 3  & 0  & 2.520 & 0.599 \\
\texttt{deepseek-chat}           & 121/150 & 80.7\%          & 5  & 16 & 2.455 & 0.667 \\
\texttt{claude-opus-4-6}         & 100/150 & 66.7\% / 100\%* & 0  & 50 & 2.990 & 0.100 \\
\bottomrule
\multicolumn{7}{l}{\footnotesize * Rate over valid non-outage runs.}
\end{tabular}
\end{table}

\subsection{Results by Vendor}
\label{sec:results_vendor}

Vendor asymmetry emerges as the strongest effect in the benchmark. As shown in Table~\ref{tab:vendor_results}, Cisco NX-OS and Arista EOS exhibit highly similar behavior, whereas Huawei VRP concentrates nearly all genuine semantic failures (11 of 12) and a disproportionate share of pipeline errors (33 of 66). The planned contrasts in Table~\ref{tab:h1_contrasts} confirm this effect: both Huawei gaps are highly significant, while Cisco and Arista remain statistically indistinguishable, fully supporting H1.

Importantly, the vendor effect amplitude reaches 17.6~p.p., which is 3.3$\times$ larger than the use-case effect amplitude (5.3~p.p.), indicating that vendor identity is the primary source of performance variation in the benchmark.

\begin{figure}[!htp]
\centering
\includegraphics[width=0.70\linewidth]{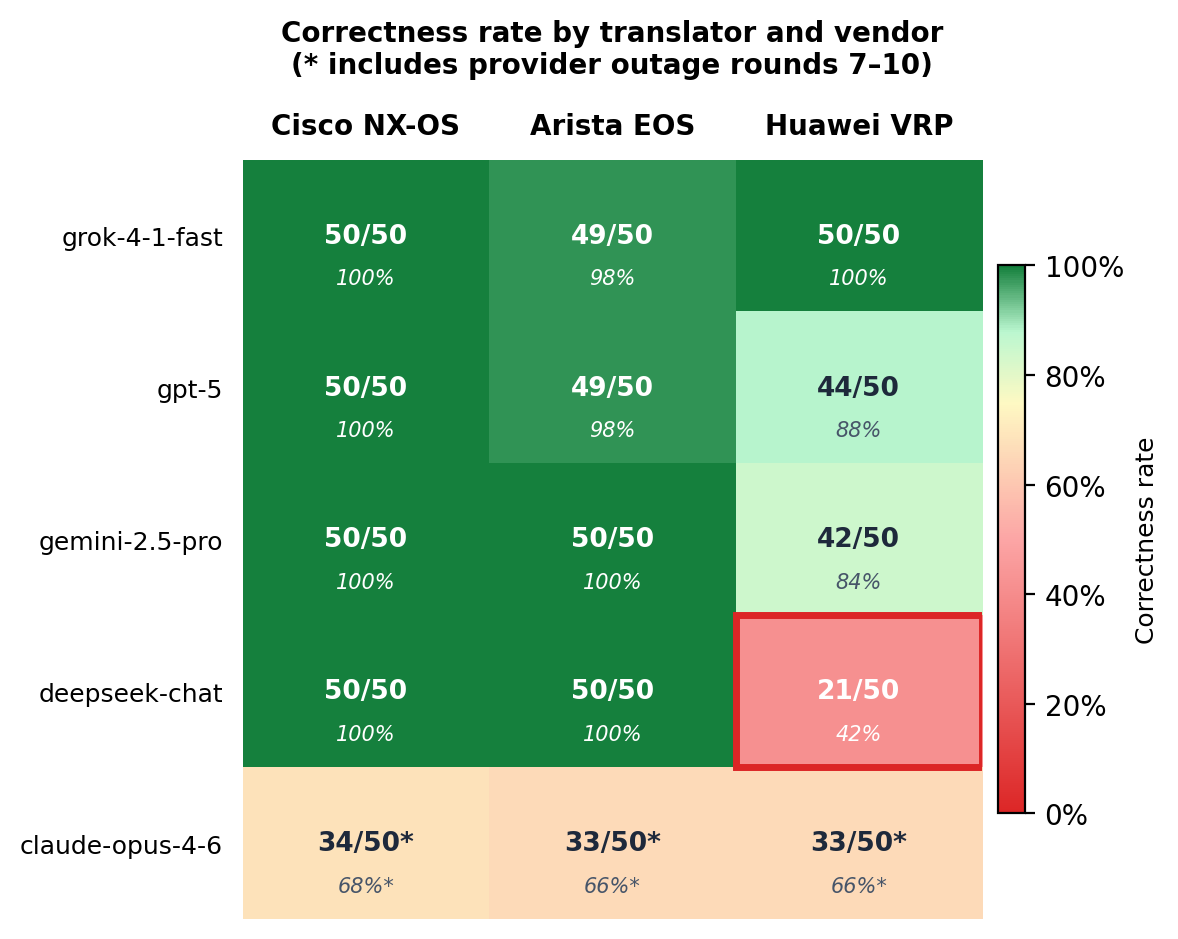}
\caption{Correctness rate by translator and vendor.}
\label{fig:heatmap}
\end{figure}

The most pronounced interaction is observed for \texttt{deepseek-chat} on Huawei VRP (Figure~\ref{fig:heatmap}). While this translator achieves 100\% correctness on Cisco and Arista, its performance collapses on Huawei, combining semantic failures in UC3 with invalid-output pipeline failures in UC4 and UC5. This pattern is consistent across rounds and would be obscured by aggregate cross-vendor summaries.

\begin{table}[!htp]
\centering
\small
\caption{Aggregate results by vendor.}
\label{tab:vendor_results}
\renewcommand{\arraystretch}{1.3}
\setlength{\tabcolsep}{4pt}
\begin{tabular}{lrrrrrr}
\toprule
\textbf{Vendor} & \textbf{Corr./Tot.} & \textbf{Rate} & \textbf{SI} & \textbf{PE} & \textbf{$\mu$ votes} \\
\midrule
Cisco NX-OS & 234/250 & 93.6\% &  0 & 16 & 2.705 \\
Arista EOS  & 231/250 & 92.4\% &  1 & 17 & 2.674 \\
Huawei VRP  & 190/250 & 76.0\% & 11 & 33 & 2.419 \\
\bottomrule
\end{tabular}
\end{table}

\begin{table}[!htp]
\centering
\small
\caption{Planned vendor contrasts for H1.}
\label{tab:h1_contrasts}
\renewcommand{\arraystretch}{1.3}
\setlength{\tabcolsep}{4pt}
\begin{tabular}{lrrl}
\toprule
\textbf{Contrast} & \textbf{$\hat{p}_1$} & \textbf{$\hat{p}_2$} & \textbf{$\Delta$ (95\% CI) / $p$} \\
\midrule
Cisco vs Huawei  & 0.936 & 0.760 & $+$0.176 [0.115, 0.237] / $<$0.001 \\
Arista vs Huawei & 0.924 & 0.760 & $+$0.164 [0.102, 0.226] / $<$0.001 \\
Cisco vs Arista  & 0.936 & 0.924 & $+$0.012 [$-$0.033, 0.057] / 0.599 \\
\bottomrule
\end{tabular}
\end{table}

\subsection{Results by Use Case}
\label{sec:results_usecase}

At the use-case level, the observed effect is weaker than the vendor dimension (Table~\ref{tab:uc_results}). When all runs are considered, the contrast between Layer~2 and SVI scenarios does not reach statistical significance; however, after excluding the \texttt{claude-opus-4-6} outage, the gap becomes nominally significant (Table~\ref{tab:h2_contrasts}). Overall, these results provide directional but modest support for H2 and indicate that the effect is sensitive to infrastructure-related disruptions.

Among the SVI scenarios, UC3 concentrates the largest number of semantic failures, largely driven by the \texttt{deepseek-chat}~$\times$~Huawei interaction discussed earlier. By contrast, UC4 appears operationally fragile, with elevated pipeline errors, but remains semantically simple. Notably, after pilot calibration, UC2 became the strongest use case in both end-to-end correctness and inter-judge agreement, suggesting that the calibration improved benchmark validity without altering its core design.

\begin{table}[!htp]
\centering
\small
\caption{Aggregate results by use case.}
\label{tab:uc_results}
\setlength{\tabcolsep}{4pt}
\begin{tabular}{llrrrrrr}
\toprule
\textbf{UC} & \textbf{Description} & \textbf{Corr./Tot.} & \textbf{Rate} & \textbf{SI} & \textbf{PE} & \textbf{$\mu$ votes} \\
\midrule
UC1 & L2 access       & 132/150 & 88.0\% & 1 & 11 & 2.583 \\
UC2 & L2 trunk + VLAN & 135/150 & 90.0\% & 3 & 10 & 2.629 \\
UC3 & SVI IPv4        & 130/150 & 86.7\% & 6 & 10 & 2.564 \\
UC4 & SVI IPv6        & 127/150 & 84.7\% & 1 & 20 & 2.638 \\
UC5 & Dual-stack SVI  & 131/150 & 87.3\% & 1 & 15 & 2.607 \\
\bottomrule
\end{tabular}
\end{table}

\begin{table}[!htp]
\centering
\small
\caption{Planned contrast for H2 (Layer~2 vs SVI).}
\label{tab:h2_contrasts}
\renewcommand{\arraystretch}{1.3}
\setlength{\tabcolsep}{4pt}
\begin{tabular}{lrrrr}
\toprule
\textbf{Contrast} & \textbf{$\hat{p}_{L2}$} & \textbf{$\hat{p}_{SVI}$} & \textbf{$\Delta$ (95\% CI)} & \textbf{$p$-value} \\
\midrule
L2 vs SVI (all runs)      & 0.890 & 0.862 & $+$0.028 [$-$0.020, 0.075] & 0.253 \\
L2 vs SVI (excl.\ outage) & 0.950 & 0.908 & $+$0.042 [$+$0.001, 0.082] & 0.044 \\
\bottomrule
\end{tabular}
\end{table}

\subsection{Stability and Reliability}
\label{sec:stability_reliability}

With ten repetitions per cell, the benchmark enables a meaningful assessment of execution stability. As a complementary analysis, we regress the Vote Stability Index (VSI) on the standard deviation of the binary correct/not-correct outcome across the 75 experimental cells. The fitted model is:
\[
\widehat{\text{VSI}} = 1.006 - 0.652\cdot\text{sd\_outcome}, \quad R^2 = 0.922,\; t(73)=-29.4,\; p < 0.001
\]
with a 95\% confidence interval for the slope of $[-0.696,\,-0.609]$.

\begin{figure}[!htp]
\centering
\includegraphics[width=0.85\linewidth]{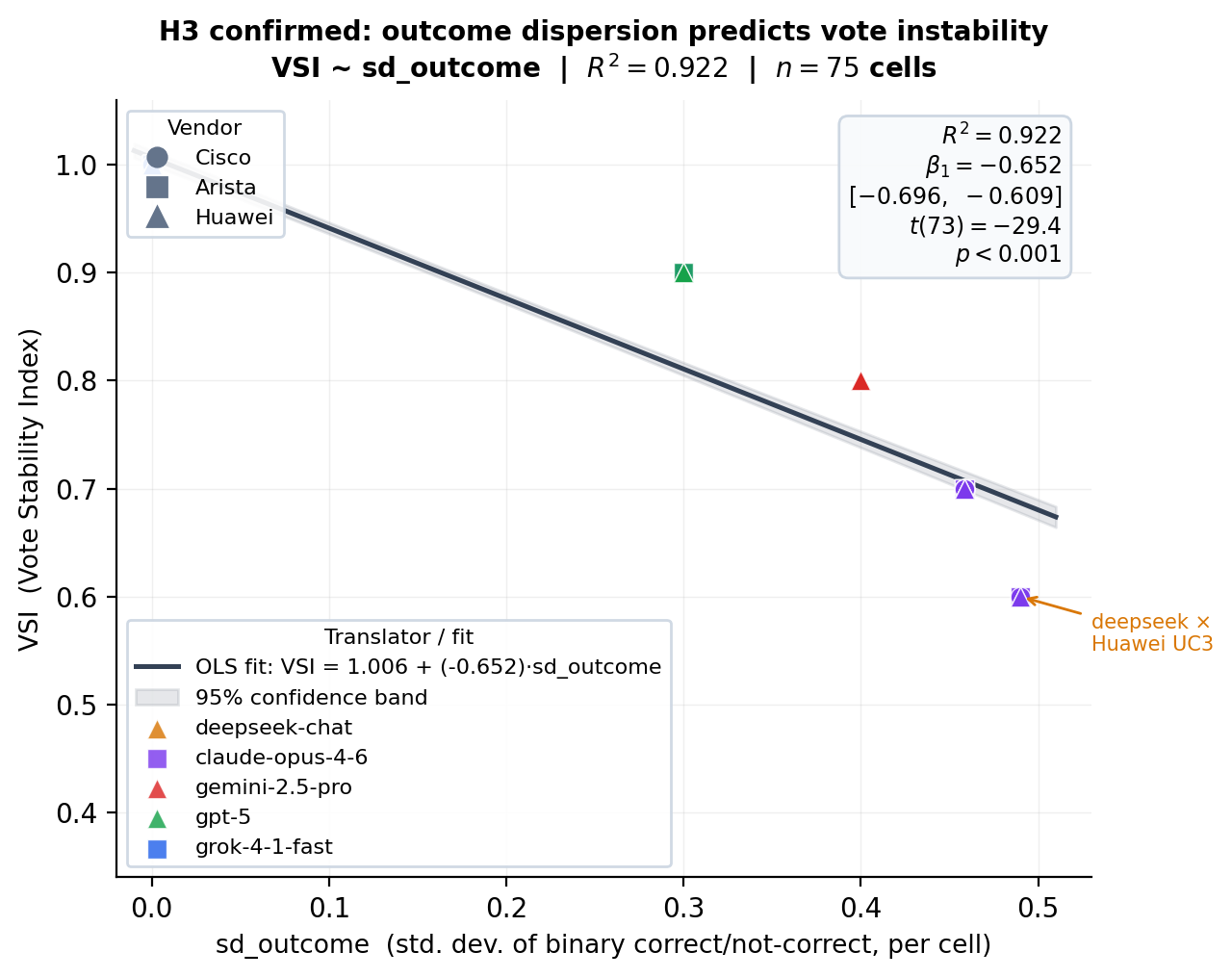}
\caption{Outcome dispersion vs.\ vote stability across 75 cells.}
\label{fig:h3_regression}
\end{figure}

The strong fit provides clear support for H3: cells with greater outcome dispersion are systematically less stable in their voting pattern (Figure~\ref{fig:h3_regression}). This relationship should be interpreted as descriptive rather than causal, since both VSI and \texttt{sd\_outcome} are derived from the same repeated outcome sequence and are therefore partially coupled by construction.

A relevant qualitative finding is that no cell simultaneously achieves perfect correctness and perfect vote stability. Even fully correct cells occasionally oscillate between 2/3 and 3/3 favorable votes, suggesting that part of the residual dispersion reflects evaluator uncertainty rather than translator instability.

\subsection{Agreement Among Judges}
\label{sec:judge_agreement}

Table~\ref{tab:kappa} reports inter-judge agreement under two complementary views. \textbf{View~A} includes all 667 runs that reached the judging stage, excluding only pipeline errors and reclassified judge faults. \textbf{View~B} further restricts the analysis to a binary-only subset of 483 runs by excluding the 183 cases in which one judge flagged a \texttt{structural} finding while the other two voted correct. These excluded cases represent categorical disagreement rather than binary correct-versus-incorrect divergence.

The two views provide a consistent picture from different statistical perspectives. In View~A, all kappa values are negative, a consequence of the prevalence paradox: because approximately 66\% of runs are unanimous, expected chance agreement becomes high, causing even moderate disagreement to push $\kappa$ below zero. This should therefore be interpreted as a property of the class distribution rather than evidence of poor judge consistency.

In View~B, after excluding structurally mixed cases and improving class balance, kappas become positive and substantially more informative. The vendor-level pattern remains stable across both views: Huawei VRP consistently concentrates the largest number of splits (83 in View~A and 38 in View~B), reinforcing the earlier finding that this vendor is intrinsically harder to evaluate consistently.

\begin{table}[!htp]
\centering
\small
\caption{Inter-judge agreement under full and binary-only views.}
\label{tab:kappa}
\renewcommand{\arraystretch}{1.3}
\setlength{\tabcolsep}{3pt}
\begin{tabular}{lrrrrrr}
\toprule
 & \multicolumn{3}{c}{\textbf{View A ($n{=}667$)}} & \multicolumn{3}{c}{\textbf{View B ($n{=}483$)}} \\
\cmidrule(lr){2-4}\cmidrule(lr){5-7}
\textbf{Stratum} & $n$ & $\kappa$ & Splits & $n$ & $\kappa$ & Splits \\
\midrule
Overall      & 667 & $-$0.077 & 225 & 483 & $+$0.197 & 41 \\
\midrule
Cisco NX-OS  & 234 & $-$0.109 &  69 & 165 & $+$1.000 &  0 \\
Arista EOS   & 232 & $-$0.104 &  73 & 162 & $+$0.244 &  3 \\
Huawei VRP   & 201 & $-$0.046 &  83 & 156 & $+$0.134 & 38 \\
\midrule
UC1          & 133 & $-$0.106 &  45 &  98 & $+$0.056 & 10 \\
UC2          & 138 & $-$0.060 &  45 &  98 & $+$0.358 &  5 \\
UC3          & 136 & $-$0.019 &  47 & 100 & $+$0.314 & 11 \\
UC4          & 128 & $-$0.100 &  42 &  91 & $+$0.148 &  5 \\
UC5          & 132 & $-$0.111 &  46 &  96 & $+$0.055 & 10 \\
\bottomrule
\end{tabular}
\end{table}

\subsection{Latency and Token Usage}
\label{sec:latency_tokens}

Operational telemetry indicates no direct correlation between correctness and latency. \texttt{grok-4-1-fast-reasoning} combines the highest semantic correctness (99.3\%) with moderate latency ($\mu{=}31.9$\,s; 7{,}834 total tokens), making it the most efficient translator overall. By contrast, \texttt{gpt-5} achieves similarly high correctness (95.3\%) at 1.9$\times$ higher mean latency ($\mu{=}60.4$\,s, p95\,=\,120.8\,s) and the highest token cost (9{,}956 tokens/run).

The higher latency of \texttt{gpt-5} is associated with longer CLI outputs (3{,}585 translator tokens versus 1{,}402--1{,}552 for the other models), which increases the downstream judge evaluation cost, combined with provider-side response-time variability. These results expose a practically relevant deployment trade-off between \emph{semantic efficiency} and \emph{semantic reliability at higher operational cost}.

For most translators, the judge layer dominates total token consumption (81--82\%), confirming that semantic verification is itself a major cost component in LLM-as-judge pipelines.

\section{Discussion}
\label{sec:discussion}

The results reveal four broader methodological and scientific implications for the evaluation of LLM-based network configuration systems.

First, semantic quality and operational reliability emerge as fundamentally orthogonal dimensions. As evidenced by \texttt{claude-opus-4-6}, a translator may remain semantically perfect on all valid judged runs while ranking substantially lower in end-to-end correctness due exclusively to infrastructure-level disruptions. This finding has direct implications for benchmark design: reporting only aggregate correctness can conflate model capability with provider-side availability and pipeline robustness. Future evaluations should therefore systematically separate semantic correctness from operational reliability, latency, and infrastructure faults.

Second, vendor identity is the dominant source of performance variation in the present benchmark. The vendor effect (17.6 p.p.) is substantially larger than the use-case effect (5.3 p.p.), and Huawei VRP concentrates nearly all genuine semantic failures and the largest share of judge disagreement. This indicates that cross-vendor generalization remains a more challenging problem than scenario complexity alone. As a consequence, aggregate rankings that omit vendor-level decomposition may be statistically misleading and operationally unsafe, particularly in heterogeneous enterprise environments.

Third, the repeated multivendor design exposes failure modes that would remain hidden under conventional single-pass evaluations. The reproducible collapse of \texttt{deepseek-chat} on Huawei VRP, despite perfect performance on Cisco and Arista, illustrates a critical benchmarking insight: average correctness can mask systematic translator-by-vendor interactions. This strongly supports the need for repeated-run experimental protocols with explicit cell-level analysis, as stochastic LLM behavior and vendor-specific semantic mappings jointly shape observed performance.

Fourth, the pilot calibration phase proved to be a substantive methodological contribution rather than merely a pre-experimental adjustment. The UC2 refinement directly improved semantic validity, leading to the highest correctness and strongest inter-judge agreement among all use cases. This demonstrates that benchmark calibration should be treated as part of the scientific method, especially in semantically evaluated systems where equivalence classes may vary across vendors.

More broadly, the results suggest that future benchmarks for LLM-based configuration translation should move beyond mean accuracy as the primary endpoint. Scientifically rigorous evaluation in this domain requires at least four explicit dimensions: semantic fidelity, infrastructure robustness, repeated-run stability, and vendor-level sensitivity. Together, these dimensions provide a substantially more faithful characterization of practical deployment risk and model behavior in real multivendor network environments.

\section{Threats to Validity}
\label{sec:threats}

A primary threat to internal validity arises from runtime conditions outside experimental control, including provider-side outages, queueing effects, and transient transport failures, which may affect latency and availability independently of semantic translation quality. We mitigate this risk through explicit separation between \texttt{pipeline error} and \texttt{semantic incorrect}, combined with randomized execution rounds to reduce temporal bias. Another internal threat concerns alignment between the intended semantic protocol and its concrete implementation in the benchmark. The pilot phase revealed overly literal judge behavior for VLAN-related constructs, which was mitigated through prompt calibration while preserving the declarative interpretation of the DSM.

A central construct-validity threat is that the benchmark measures semantic adherence under a fixed judging protocol rather than operational correctness in a live network. Consequently, the claims in this paper are limited to semantic consistency with the DSM and do not directly imply production-level service correctness. An additional concern stems from the use of LLM-based judges instead of formal verifiers, as their outputs may be affected by prompt sensitivity and representational bias. This threat is mitigated through a fixed heterogeneous judge panel, preservation of all evaluation artifacts, and explicit inter-judge agreement analysis.

External validity is necessarily bounded by the experimental scope. Although the benchmark spans three vendors and five representative use cases, important configuration domains, such as routing policies, ACLs, QoS, and VPN services, remain outside the present study. Moreover, because vendor effects dominate use-case effects in the current benchmark, different vendor compositions may lead to substantially different aggregate rankings. The results should therefore be interpreted as benchmark-specific rather than vendor-neutral assessments of general LLM capability.

A final threat concerns long-term reproducibility and benchmark drift. Because the study relies on cloud-hosted models referenced by provider-facing identifiers, the underlying implementations may evolve without notice. Although we preserve execution artifacts, model identifiers, outputs, failure labels, and telemetry, long-term reproducibility may still be affected by provider-side model updates.

\section{Conclusions and Future Work}
\label{sec:conclusion}

This paper presented a reproducible semantic benchmark for systematically comparing cloud LLM translators in the task of converting structured network intents, expressed as Desired State Models (DSMs), into vendor-specific CLI. The benchmark spans five translators, three vendors, five representative use cases, and ten repetitions per experimental cell, totaling 750 executions under a controlled protocol with fixed judges, an explicit failure taxonomy, and full artifact preservation.

The results provide strong evidence that scientifically rigorous evaluation in this domain requires a multidimensional perspective. H1 was fully confirmed, showing that vendor effects dominate use-case effects and constitute the primary source of performance variation in the benchmark. H3 received strong descriptive support, with outcome dispersion strongly associated with vote instability across the 75 cells, although this relationship should be interpreted cautiously due to partial structural coupling between the variables. H2 received only directional and modest support.

More broadly, the study demonstrates that semantic quality and operational reliability are orthogonal dimensions, and that aggregate correctness alone can hide critical vendor-specific failure modes, as evidenced by the reproducible collapse of \texttt{deepseek-chat} on Huawei VRP. These findings reinforce that future evaluations of intent-to-CLI translators should explicitly separate semantic correctness from infrastructure robustness, decompose results by vendor, and incorporate repeated executions as a first-class component of the experimental protocol.

Finally, future work includes expanding the benchmark to additional vendors, configuration domains, and more complex operational scenarios, such as routing policies, ACLs, QoS, and VPN services. We also plan to investigate hybrid evaluation pipelines that combine LLM-based judging with formal verification and execution-based validation, as well as the inclusion of open-weight and domain-adapted translators. A particularly important direction is the longitudinal re-execution of fixed benchmark manifests to quantify benchmark drift and provider-side model evolution over time.

\bibliographystyle{plain}
\bibliography{references}

@inproceedings{lin2025agentic,
  title={Agentic Framework for Natural Language to Network Configuration Translation},
  author={Lin, X. and others},
  booktitle={Proceedings of CNSM},
  year={2025}
}

@inproceedings{net2dllm2025,
 author = {Jerônimo Menezes and Leonardo Bitzki and Diego Kreutz},
 title = { {Net2d-LLM}: Translating Structured Network Intents into {CLI} using {LLMs} with Execution in a Network Digital Twin},
 booktitle = {Anais da XXII Escola Regional de Redes de Computadores},
 location = {Porto Alegre/RS},
 year = {2025},
 issn = {0000-0000},
 pages = {54--60},
 publisher = {SBC},
 xaddress = {Porto Alegre, RS, Brasil},
 doi = {10.5753/errc.2025.17772},
 url = {https://sol.sbc.org.br/index.php/errc/article/view/39181}
}

@inproceedings{netconfeval2025,
  title={Netconfeval: Can {LLMs} facilitate network configuration?},
  author={Wang, Changjie and Scazzariello, Mariano and Farshin, Alireza and Ferlin, Simone and Kosti{\'c}, Dejan and Chiesa, Marco},
  booktitle={Proceedings of the ACM on Networking},
  pages={1--25},
  year={2024},
  publisher={ACM New York, NY, USA}
}

@inproceedings{peeringllmbench2025,
  title={{PeeringLLM-Bench}: Evaluating {LLMs} for {BGP} Configuration Tasks},
  author={Mendoza, John Robert and Ocampo, Roel},
  booktitle={Proceedings of the 20th Asian Internet Engineering Conference},
  xpages={78--86},
  year={2025}
}

@inproceedings{netllmbench2025,
  title={Netllmbench: A benchmark framework for large language models in network configuration tasks},
  author={Aykurt, Kaan and Blenk, Andreas and Kellerer, Wolfgang},
  booktitle={IEEE Conference on Network Function Virtualization and Software Defined Networks (NFV-SDN)},
  xpages={1--6},
  year={2024},
  organization={IEEE}
}

@inproceedings{inta2025,
  title={Inta: Intent-based translation for network configuration with llm agents},
  author={Wei, Yunze and Xie, Xiaohui and Hu, Tianshuo and Zuo, Yiwei and Chen, Xinyi and Chi, Kaiwen and Cui, Yong},
  booktitle={2025 IEEE 33rd International Conference on Network Protocols (ICNP)},
  pages={1--16},
  year={2025},
  organization={IEEE}
}

@article{liu2024large,
  title={Large Language Models for Networking: Workflow, Advances, and Challenges},
  author={Liu, Chang and Xie, Xiaohui and Zhang, Xinggong and Cui, Yong},
  journal={IEEE Network},
  volume={39},
  number={5},
  pages={165--172},
  year={2024},
  publisher={IEEE}
}

@article{boateng2025survey,
  title={A survey on large language models for communication, network, and service management: Application insights, challenges, and future directions},
  author={Boateng, Gordon Owusu and Sami, Hani and Alagha, Ahmed and Elmekki, Hanae and Hammoud, Ahmad and Mizouni, Rabeb and Mourad, Azzam and Otrok, Hadi and Bentahar, Jamal and Muhaidat, Sami and others},
  journal={IEEE Communications Surveys \& Tutorials},
  year={2025},
  publisher={IEEE}
}

@article{hong2025comprehensive,
  title={A Comprehensive Survey on LLM-Based Network Management and Operations},
  author={Hong, Jibum and Tu, Nguyen Van and Hong, James Won-Ki},
  journal={International Journal of Network Management},
  volume={35},
  number={6},
  pages={e70029},
  year={2025},
  publisher={Wiley Online Library}
}

@article{long2025survey,
  title={A survey on intelligent network operations and performance optimization based on large language models},
  author={Long, Sifan and Tan, Jingjing and Mao, Bomin and Tang, Fengxiao and Li, Yangfan and Zhao, Ming and Kato, Nei},
  journal={IEEE Communications Surveys \& Tutorials},
  volume={27},
  number={6},
  pages={3915--3949},
  year={2025},
  publisher={IEEE}
}

@inproceedings{tageldien2025large,
  title={Large language models in intent-based networking: a comprehensive survey across the intent lifecycle},
  author={Tageldien, Marwa and Selim, Bassant and Sboui, Lokman},
  booktitle={ITC-Egypt},
  pages={810--817},
  year={2025},
  organization={IEEE}
}

@misc{wei_inta_2025,
  title         = {{INTA}: Intent-Based Translation for Network Configuration with {LLM} Agents},
  author        = {Wei, Yunze and Xie, Xiaohui and Hu, Tianshuo and Zuo, Yiwei and Chen, Xinyi and Chi, Kaiwen and Cui, Yong},
  year          = {2025},
  howpublished  = {arXiv preprint},
  eprint        = {2501.08760},
  archivePrefix = {arXiv},
  primaryClass  = {cs.NI},
  doi           = {10.48550/ARXIV.2501.08760},
  url           = {https://arxiv.org/abs/2501.08760}
}

@article{llm-powered,
author = {Wang, Jingyu and He, Bo and Zhao, Jinyu and Xuan, Yixin and Sun, Haifeng and Qi, Qi and Liang, Junzhe and Zhuang, Zirui and Liao, Jianxin},
year = {2026},
month = {01},
pages = {1-1},
title = {LLM-Powered Intent-Driven Configuration Generation for Multi-Vendor Networks},
volume = {PP},
journal = {IEEE Transactions on Network and Service Management},
doi = {10.1109/TNSM.2026.3675409}
}

@INPROCEEDINGS{11349540,
  author={Raptis, Nikos and Adhane, Gereziher and Fonseca, João Pedro and Ramantas, Kostas and Verikoukis, Christos},
  booktitle={2025 IEEE Conference on Network Function Virtualization and Software-Defined Networking (NFV-SDN)}, 
  title={ARGVI: Adaptive Routing, Generation, and Validation of Intents for Intent-Driven Management}, 
  year={2025},
  volume={},
  number={},
  pages={1-6},
  doi={10.1109/NFV-SDN66355.2025.11349540}}

\end{document}